\documentclass[conference]{IEEEtran}
\IEEEoverridecommandlockouts
\usepackage{cite}
\usepackage{amsmath,amssymb,amsfonts}
\usepackage{algorithmic}
\usepackage{graphicx}
\usepackage{textcomp}
\usepackage{xcolor}
\usepackage{balance}
\usepackage{diagbox}

\renewcommand{\baselinestretch}{0.855}

\newcommand{\system}{\textbf{CWD}}

\def\BibTeX{{\rm B\kern-.05em{\sc i\kern-.025em b}\kern-.08em
    T\kern-.1667em\lower.7ex\hbox{E}\kern-.125emX}}

\makeatletter
\newcommand{\linebreakand}{%
  \end{@IEEEauthorhalign}
  \hfill\mbox{}\par
  \mbox{}\hfill\begin{@IEEEauthorhalign}
}
\makeatother

\begin{document}

\title{{CWD: A Machine Learning based Approach to Detect Unknown Cloud Workloads}
{\footnotesize \textsuperscript{}}
\thanks{}
}

\author{\IEEEauthorblockN{1\textsuperscript{st} Mohammad Hossain }
\IEEEauthorblockA{\textit{Intel Corporation} \\
mohammad1.hossain@intel.com}
\and
\IEEEauthorblockN{2\textsuperscript{nd} Derssie Mebratu}
\IEEEauthorblockA{\textit{Intel Corporation} \\
derssie.d.mebratu@intel.com}
\and
\IEEEauthorblockN{2\textsuperscript{rd} Niranjan Hasabnis}
\IEEEauthorblockA{\textit{Intel Lab} \\
niranjan.hasabnis@intel.com}  
\and
\linebreakand 
\IEEEauthorblockN{3\textsuperscript{th}  Jun Jin }
\IEEEauthorblockA{\textit{Intel Corporation}\\
jun.i.jin@intel.com}

\and
\IEEEauthorblockN{4\textsuperscript{th} Gaurav Chaudhary }
\IEEEauthorblockA{\textit{Intel Corporation}\\
gaurav1.chaudhary@intel.com}
\and
\IEEEauthorblockN{4\textsuperscript{th} Noah Shen }
\IEEEauthorblockA{\textit{Intel Corporation}\\
noah.shen@intel.com}

}

\maketitle

\begin{abstract}

Workloads in modern cloud data centers are becoming increasingly complex. The
number of workloads running in cloud data centers has been growing exponentially
for the last few years, and cloud service providers (CSP) have been supporting
on-demand services in real-time. Realizing the growing complexity of cloud
environment and cloud workloads, hardware vendors such as Intel and AMD are
increasingly introducing cloud-specific workload acceleration features in their
CPU platforms. These features are typically targeted towards popular and
commonly-used cloud workloads. Nonetheless, uncommon, customer-specific
workloads (unknown workloads), if their characteristics are different from
common workloads (known workloads), may not realize the potential of the
underlying platform. To address this problem of realizing the full potential of
the underlying platform, we develop a machine learning based technique to
characterize, profile and predict workloads running in the cloud environment.
Experimental evaluation of our technique demonstrates good prediction
performance. We also develop techniques to analyze the performance of the model
in a standalone manner.




\end{abstract}

\begin{IEEEkeywords}
machine learning, gaussian mixture model, dynamic sliding window, scan statistical analysis and simulation

\end{IEEEkeywords}

\section{Introduction}

Workloads in modern cloud data centers are becoming increasingly complex.
Moreover, the number of such workloads is growing at an unprecedented rate. Data
Center Industry Survey Results, published by Uptime Institute in
2019~\cite{dcuptime} found that although the majority of IT workloads are still
on enterprise data centers, data center workloads are becoming both complex and
extensive.  More importantly, it also found that cloud capacity is growing at a
faster rate than the capacity of the enterprise data center and that IT
workloads continue to move to the cloud. For instance, software-development
paradigms are transitioning from monolithic applications to smaller,
finer-grained, and distributed services (microservices), and cloud service
providers (CSP) have started offering \emph{serverless computing} platforms that
bill resources at finer granularities (e.g., Function-as-a-services (FaaS)).
For example, NetFlix and Salesforce are increasingly adopting cloud-native
principles (e.g., \emph{cloud-native functions}, which is a type of
microservice) and architectural design principles to support on-demand services.
New specific-purpose models, such as FaaS, add to the complexity of cloud
environments that are already supporting more general-purpose models such as
infrastructure-as-a-service (IaaS), software-as-a-service (SaaS),
platform-as-a-service (PaaS). In addition to workload complexity, even the
deployment models are becoming more complex.  Specifically, workload
consolidation --- co-locating different workloads to drive up utilization and
improve efficiently (e.g., energy efficiency) --- is a common model in most of
the public clouds now. At the hardware level, resource disaggregation is a
popular approach to avoid resource stranding in data centers.

Realizing the growing trend of workload migration to cloud and the need of
accelerating growing diversity of cloud workloads, hardware vendors such as
Intel and AMD already offer several performance-oriented features in their
cloud-specific platforms. For instance, Intel's recently launched 3rd generation
Xeon Scalable processors are routinely used in cloud environments, and they
offer several features (e.g., Intel Total Memory Encryption that protects data
and VMs by encrypting their physical memory, Intel Crypto Acceleration that
accelerates commonly-used hash functions and encryption protocols, etc.) that
accelerate common cloud workloads. Such platform features, when combined with
software optimizations, accelerate commonly-used, popular cloud workloads and
microservices.

One of the common problems faced by hardware vendors in cloud environments is
the inability to analyze performance bottlenecks and resource utilization of
cloud workloads. This is partly because of the fact that the hardware vendors
have limited access in cloud environments; typically, they have access limited
access at the hardware level, let alone the software stack. Lack of such
analysis makes it difficult to evaluate impact of different hardware features on
workload performance.  Operating within such a constrained environment, in this
paper, we attempt to answer a question of \textit{if it is possible to identify
the workload running on a given system in cloud}. To answer this question, we
develop a machine learning (ML) based system, named {\system} (Cloud Workload
Detection), that utilizes hardware performance counters to develop workload
\textit{fingerprints} and then employs an efficient approach to compare
fingerprints to detect unknown workloads. Intuitively, fingerprint of a workload
is its unique signature that distinguishes the execution of that workload from
others. Our experimental results demonstrate efficacy of {\system} in detecting
unknown workloads with a high degree of accuracy.

Although, we do not discuss possible applications of detecting unknown cloud
workloads in this paper, it is not difficult to imagine a number of
applications. For instance, one can propose possible optimizations to the
unknown workload, if the performance of that workload is lower than its expected
performance on the hardware platform (e.g., a crypto workload is not using
crypto acceleration engine in the hardware). Another application could be of
enabling privacy-preserving analysis of cloud workloads --- the hardware vendors
could apply the approach described in this paper to identify a proxy workload
that has similar fingerprint as that of a proprietary cloud workload and then
analyze the proxy workload instead of the actual workload.

\section{Related work}
As a new cloud workload continuously emerge, designing and deploying a reliable physical computer hardwares in a data center are extremely vital. The hardware includes storage, networking,  processor, memory and etc. Furthermore, without evaluating, analyzing and identifying the characteristics and profile of a workload running in a cloud, it is very difficult  to optimize and improve the performance of that workload. To solve this problem, many researchers have proposed various methods.  Khanna et al. developed \cite{Khanna2013} a workload phase forecasting method to evaluate and analyze the run-time behavior of workloads in a cloud. The results obtained that the new model of workload phase shows that 98\%  accurate in phase identification and 97.15\% accurate in forecasting the compute demands. In addition, Jandaghi et al. \cite{Jandaghi2018} studied workload profile and resource usage prediction technique to detect a workload in Apache Spark framework using a gaussian mixture model. Furthermore, Elnaffar et al. \cite{Elnaffar2002} introduced a new model that characterizing a workload. As the author noted that the method helps to evaluate and predict workload performance.  Bhattacharyya et al. \cite{Bhattacharyya2020} proposed a performance modeling technique for cloud workload to evaluate and predict workloads performance by identifying automatically workload phases. The results showed that this method can predict the performance of applications with up to 95\% accuracy for previously unseen input configurations at less than 5\% overhead. Moreover, \cite{Bhattacharyya2016} introduced for automatic phase detection and characterization applications running in a cloud.  The results achieved detecting a phases upto 98.2\% accuracy with an average detection delay of less than 0.01 seconds. Besides, Sherwood et al. \cite{Sherwood2003} introduced a unified profiling architecture to capture, classify, and predict phase-based program behavior of workloads. Furthermore, Calheiros et al. \cite{Calheiros2015} presented a cloud workload prediction model for cloud service providers using the autoregressive integrated moving average (ARIMA). As Desprez et al. \cite{Desprez2010} has proposed that a workload prediction based on a similar past occurrences of the current short-term workload history; present an overall evaluation of this approach and usefulness for enabling efficient auto-scaling of cloud resources. 

We develop a machine learning based technique to characterize, profile and predict workloads running in a cloud environment. Experimental evaluation of our technique demonstrates good prediction performance. We also develop techniques to analyze the performance of the model in a standalone manner.

The rest of the paper is laid out as follows.The first section describes the background information, prior art, and related work done in the domain of workload characterization and analysis, workload profile methodologies, and workload phase detection methods. The following section includes a theoretical description of the machine learning methodology and phase classification, phase evaluation, correlation changepoint detection, and probabilistic and statistical models for phase-different workload. The experiment section includes our setup and analysis of observed result data from real dataset. The fourth section proposes the result and analysis. Finally, the conclusion and future work are discussed in the last section.

\section{Algorithms}
 
In this paper, we combine three algorithms to predict unknown workload running
in a cloud: a) change point detection,  b) Euclidean distance and c) dynamic
time warping (DTW)

\subsection{Change Point Detection}

Change point detection technique can be used to locate the transition between
phases or states in a signal or time-series data\cite{Truong2019}. The
occurrence of the state depends on some underlying dynamic change on the system.
For example, when a noise is added into a signal or time series data (hardware
and software performance counters), it causes an abrupt change to the sequence
of a signal or time series data. The detection of abrupt change(change points)
in a signal or time series data is a well studied problem, several authors have
proposed several techniques such as outlier detection, relative density ratio
estimation and etc.  In this research, we use a multivariate change point
detection method to identify the state transitions of a time series data for
hardware and software performance counters such as  L1 cache misses, LLC cache
misses, CPU utilization, iTLB misses, etc.

\subsubsection{Problem Formulation}

we assume that the sequence of observations can be expressed as $y_{1:n}$$=(y_{1}$, $y_{2,\cdots,}$$y_{n}$)$^{T}$ denote $n$ data samples observed. Each data samples lies
in $R^{d}$ and $y_{n}$$\epsilon$ $\mathbb{R^{d}}$,
$\forall$ $n$$=1,2,\cdots,N.$ Note that $d$ is a dimensional space
records hardware and software counter data. $y_{i:j}$ represents
the data between time indices $i$ and $j$, where $(y_{i}$,$y_{i+1,\cdots,}$$y_{j}$)$^{T}$.
In this case, there is some $M$ change points in time series 
data points sequence, denoted in increasing order based on the following
indices such as $t_{1}$,...,$t_{m}$ and let $t_{0}$$=0$ and $t_{M+1}$$=N$.
To identify the transition between phases, we apply
Bayesian Change Point Detection(BCPD) algorithm\cite{Malladi2013} which is based 
on Bayes' rule \cite{Josang2016} that describes the probability of a prior distribution and 
likelihood distribution  and computes the posterior distribution $p\left(x_{t}\mid y_{1:t}\right)$ as shown equation (5). 
\begin{equation}
p\left(x_{t}\mid y_{1:t}\right)=\frac{p(y_{1:t}\mid x_{t})\centerdot p(x_{t})}{p(y_{1:t})}
\end{equation}

Where $p(x_{t})$ is a vector and prior observation of hardware and
software counters of time series data and $x_{t}$ is the run length of the change point detection algorithm.
 $p(y_{1:t}\mid x_{t})$
is a vector and the current observation and the joint likelihood distribution
of hardware and software counters of time series data. The joint likelihood distribution is computed based on each new observation of the counters. We can express the joint distribution over run length and observed data recursively.

\begin{align}
p\left(x_{t}\mid y_{1:t}\right) & =\sum p(x_{t},x_{t-1},y_{1:t})\\
 & =\sum p(x_{t},y_{t}\mid x_{t-1,}y_{1:t-1})p(x_{t-1},y_{1:t-1})\\
 & =\sum p(x_{t},x_{t-1})p(y_{t}\mid x_{t-1,}y_{t}^{(x)})p(x_{t-1},x_{1:t-1})
\end{align}
\\
As equ. (4) shows that based on the newly observed and prior data point, the posterior distribution predicts the new change point of a multivariate time series data of hardware and software performance counters.

Moreover, the advantage of using BCPD is that evaluating change point in a multivariate time-series data is better than the other change point detection techniques such as Nonparametric Change Point Detection or Kernel Change Point Detections \cite{bcpd}. 

\subsection{Euclidean Distance}

Euclidean distance is a well-known distance measurement between two time series
data have the same length. In this research, as shown in equation (5),  we use the
Euclidean distance to compare the similarity between known workload to unknown
workload:

\begin{equation}
d(x_{i},x_{j})=\sqrt{\sum_{r=1}^{n}(x_{i}-x_{j})^{2}}
\end{equation}

Where known workload vector space is $x_i$ $=(c_{1,}c_{2,\cdots,}c_{n})$ and unknown workload vector space is $x_j$ $=(b_{1,}b_{2,\cdots,}b_{n})$. If there is a smaller distance between $(x_{i},x_{j})$, the two workloads are similar.

\subsection{Dynamic Time Warping (DTW)}

DTW is also used to measure the similarity between two time series data. Unlike
Euclidean distance, DTW measures the distance between two data points that do not
have equal lengths. Assume, there are two time series and represent with known
workload $P$ and unknown workload $Q$ and have $n$ by $m$ matrix and is
constructed based on ($i^{th}$, $j^{th}$) and represented by a warping path $W$
$=(w_{1,}w_{2,\cdots,}w_{n})$. $P$ $=(p_{1,}p_{2,\cdots,}p_{n})$ and $Q$
$=(q_{1,}q_{2,\cdots,}q_{n})$. Thus, known and unknown workloads time series
measurement between $P$ and $Q$:

\begin{equation}
DTW(P,Q)=\underset{W=w_{1},...,w_{k}}{arg\ min}\sum_{k=1,w_{k}=(i,j)}^{k}\left( p_{i}-q_{i} \right)^{2}
\end{equation}

%
%

\section{Problem Formulation}

As mentioned before, in this paper we aim to identify workloads running on
hardware (in cloud, on premise, or on a bare metal), without having any access
to the workloads. More specifically, we aim to identify workloads running on
hardware by using hardware-level features and without having any knowledge of
and access to the software environment (e.g., operating system, libraries, etc.)
Given this constrained environment, the only ``view'' that the hardware can
offer is via workload execution. In particular, we can leverage processor-level
telemetry information that offers insights into workload's interactions with
CPU, memory, network, and I/O subsystems, etc.

It is important, however, to note that the telemetry information for a program
execution could be specific to a particular program input. In other words, the
telemetry information could be different for a different program input. This is
because different inputs could follow different program execution paths.
Similarly, different execution environments (such as opearting system version,
configuration of an existing operating system, etc.) could produce different
telemetry data. Telemetry data obtained from workload execution is thus
dependant on all of such factors. We conceptually model this phenomenon as a
mapping function $f$:

\[ t = f(w, i, \Sigma) \]

Where $t$ is the telemetry data for the execution of a workload $w$, and $w$ is
given $i$ as input and has $\Sigma$ as the execution environment. Note that the
definition of $\Sigma$ is not provided as it is vast and encompasses
fine-grained details --- all the way from software to hardware, such as hardware
configuration, versions of software components, etc., --- that are required to
deterministically product telemetry data $t$ for workload $w$, when given input
$i$.

The problem of identifying unknown workload, defined via its execution (and
represented as telemetry information $t'$), is then finding $w'$, $i'$, and
$\Sigma'$, such that:

\[ t' = f(w', i', \Sigma') \]

Note that the function $f$ is unknown, and in a sense, that is what our ML model
learns by synthesizing it from several examples of its input/output mappings.

Although, it is easy to model this problem conceptually, note that it is a
complex problem as $\Sigma$ is a set of vast number of parameters. We simplify
the problem to make it tractable by keeping some of them fixed: Specifically, we
assume that the unknown workload is run on the same hardware platform and the
system as those used for the known workloads. That way parameters related to
hardware setup need not be part of $\Sigma$.

As a side note, realize that the fundamental assumption in this modelling is
that the workload execution is deterministic; if a workload has
non-deterministic behavior because of bugs or other reasons, then the current
modelling approach does not support identifying such workloads with high degree
of accuracy.

\section{System Design}

\begin{figure}[tbp]
\centering
\includegraphics[width=0.5\textwidth]{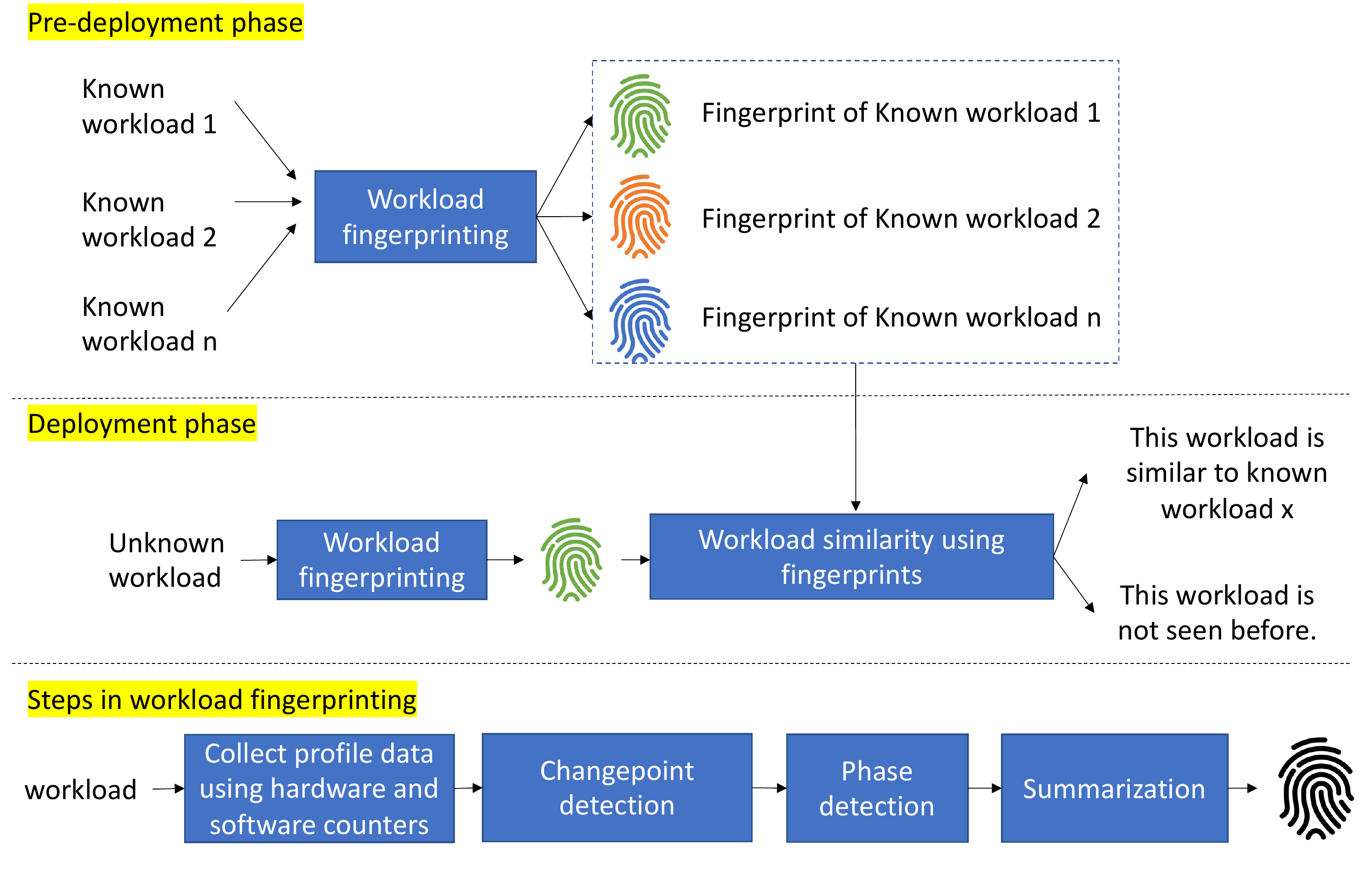}
\caption{Flow chart: Unknown workload prediction}
\label{fig:system_design}
\end{figure}

Figure~\ref{fig:system_design} shows the overall design of our system. At a
high-level, our system can be broken into two major phases: a training phase and
an inference phase. The purpose of the training phase is essentially to learn
the ``fingerprints'' of the known workloads, while the purpose of the inference
phase is to detect an unknown workload by obtaining its ``fingerprints'' and
comparing them to the fingerprints of the known workloads. The approach of
comparing the fingerprints relies on the notion of workload \textit{similarity}.

The steps performed by our system to obtain fingerprints of a workload are as
follows: \emph{(i)} executing the workload on a target system while collecting
hardware and (allowed) software counters while running the workload, \emph{(ii)}
detecting changes in the workload phases by using Bayesian change point
detection (BCPD), \emph{(iii)} summarizing workload phases, and \emph{(v)}
generating fingerprints using summarized phases. Given a set of workloads to be
used for training phase, we perform these steps on all the workloads
sequentially and collect fingerprints for each of them.


\subsection{Telemetry Data Collection}

Given a workload along with its test input, we first execute the workload and
collect its profiling information by using certain hardware and software
counters. In a nutshell, given a set of $n$ counters and time of $t$ seconds for
which the workload is run, the output of this phase is a 2D matrix of size
$n{\times}t$ (assuming that the counters were sampled for every one second). In
other words, the output is a set of $n$ discrete time series. Note that the absolute
values of the counters would have different magnitudes. Hence, each set of
time series data is normalized using the following equation:

\begin{equation}
\label{eq:normalize}
y = \frac{|x - \mu|}{\sigma}
\end{equation}
where $y$, $x$, $\mu$ and $\sigma$ are a normalized value, a raw data value, the
arithmetic mean and the standard deviation of time series data, respectively.

\begin{figure}[!htbp]
\centerline{\includegraphics[width=1.2\linewidth]{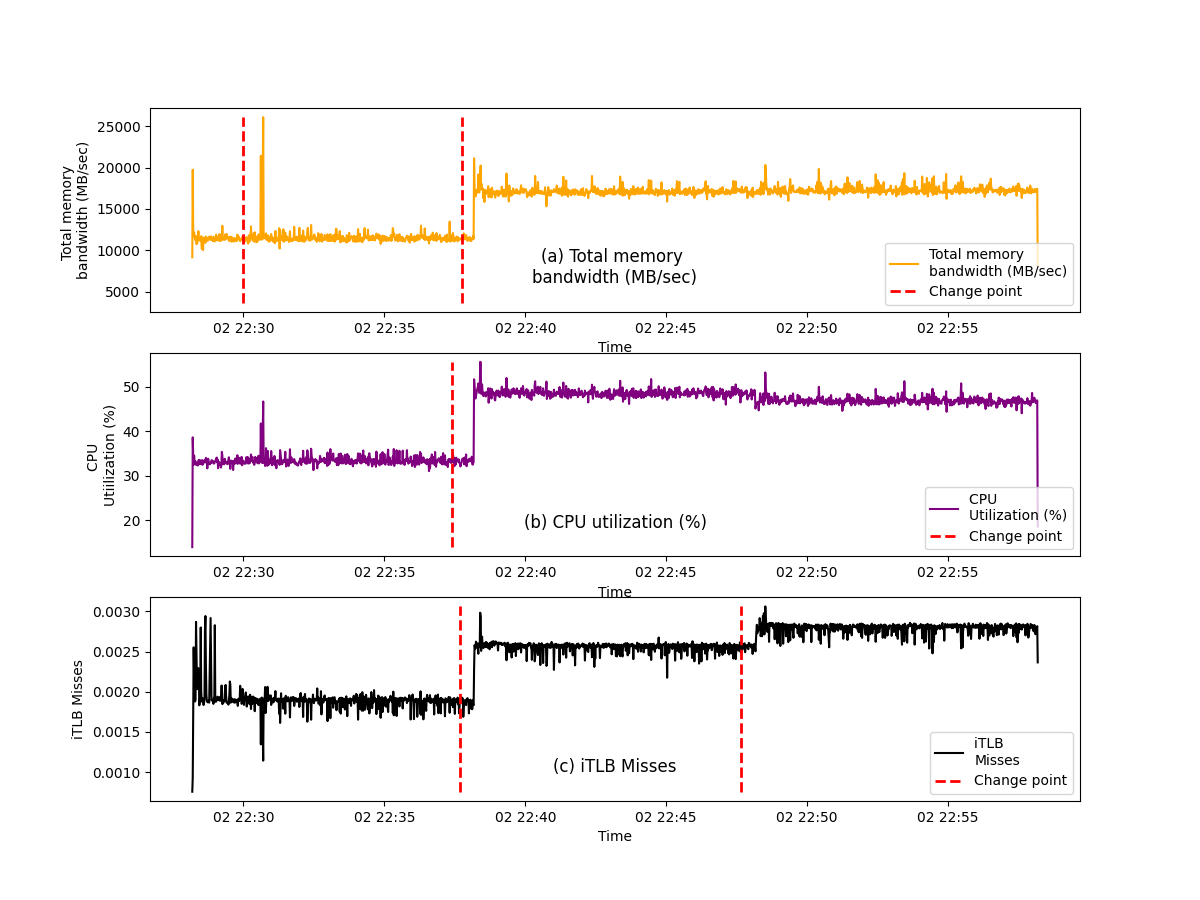}}
\caption{Phase identification using BCPD algorithm}
\label{fig:change_point_detection}
\end{figure}

\subsection{Change Point Detection}

In this step, we apply the Bayesian change point detection technique (BCPD) to
individual counter's normalized time series data to locate the changepoints,
which essentially represent the endpoints of phases. Intuitively, we breakdown
time series for a counter into components (called \textit{phases}) that are
``similar'' in nature (statistically). 
For example, Figure~\ref{fig:change_point_detection} shows the change points (in
red dotted line) identified by the BCPD technique for the three different time
series of total memory bandwidth,  CPU utilization (\%), and iTLB misses for
MongoDB-ycsb workload. Note that different time series could have different
number of changepoints. For instance, time series of total memory bandwidth and
iTLB misses have two changepoints, while that of CPU utilization has only one
changepoint.


\subsection{Workload Fingerprinting}

The aim of this step is to generate workload-specific (in other words, unique)
signature that allows us to identify that workload. We call such a unique
signature a \textit{fingerprint}. Concretely, we define workload fingerprint as
the summary of all the phases identified in all the time series data for that
workload.  In this step, we use statistical measures to summarize every phase.
Specifically, for each phase, we calculate the mean, median, variance, augmented
Dickey-Fuller test statistics, and the distribution of the phase.  If a time
series has $p$ phases and $k$ is the number statics measured per phase, then the
summary of that time series would be a matrix of size $p{\times}k$.  However,
the number of phases identified in different time series may not be same.  As a
result, the time series summary matrix could vary in size from counter to
counter based on the number of phases ($p$). We alleviate this issue by applying
zero padding to the summary matrices based on the maximum number of phases
across all counters for a given workload.  For example, if a workload profile
data has $n$ counters, and $q$ is the highest number of phases identified across
all the counters, then the workload fingerprint would be a table of size
$n{\times}q{\times}k$, where $q \le p$.

\subsection{Workload Similarity}
\label{workload_similarity}

This step of workload similarity is aimed at evaluating similarity of two
workloads, specified using their fingerprints obtained in the previous step.
Conceptually, two workloads would be similar (denoted as $\sim$) if (1) they
both have same number of phases and (2) all the phases of both the workloads are
mutually similar (denoted by $\approx$), meaning phase $i$ of one workload is
similar to phase $i$ of the other workload.  Formally, if we represent
fingerprint of one workload as $(t_1, t_2, .., t_m)$, where $m$ is the number of
phases of that workload and $t$ is the fingerprint of a phase, and the
fingerprint of another workload is $(t'_1, t'_2, .., t'_n)$, where $n$ is the
number of phases of the second workload and $t'$ is the fingerprint of its phase,
then $(t_1, .., t_m) \sim (t'_1, .., t'_n))$ if:

\[ \forall i \in [1, n] \mid t_i \approx t'_i \quad \textit{if } m = n \]

This definition of workload similarity, however, does not consider workloads
having different number of phases. A simple approach to handle workloads having
different number of phases could be to run them from start to end. It,
nonetheless, restricts efficacy of detecting unknown workloads that are not run
to completion (in other words, a small window into their execution). We
consequently extend the earlier definition of workload similarity to consider
this scenario. Specifically, what we need to check for is whether the workload
having smaller number of phases is a portion of the workload having larger
number of phases (similar to substring matching problem). Formally, we modify
the above formulation as $(t_1, .., t_m) \sim (t'_1, .., t'_n))$ if:

\[
  \begin{cases}
    \forall i \in [1, m], \exists j \in \mathbb{N} \mid t_i \approx t'_{i+j} & \quad \text{if } m < n \\
    \forall i \in [1, n], \exists j \in \mathbb{N} \mid t_{i+j} \approx t'_i & \quad \text{if } m >= n
  \end{cases}
\]

In the above equation, $j$ essentially represents an \textit{offset} into the
phases of the workload having larger number of phases. In other words, the
workload having smaller number of phases could start at any offset within the
workload having larger number of phases.

In this paper, we use Euclidean distance and dynamic time warping (DTW)
technique with Manhattan distance to implement phase-level similarity function
($\approx$).


The similarity measurement between the phases of the same workload provides a
threshold value to identify similar workloads. For this study, the threshold is
defined as \texttt{mean of phase-level similarity values} $\pm$ \texttt{standard
deviation}. If an unknown workload's similarity measurement falls within the
threshold values of a particular workload, then we can say the unknown workload
is similar to  that particular workload.  To validate the threshold value, we
randomly split the data set 70\%-30\% ratio to create train and validation data
set.  Then, we establish the threshold value from the mean and standard
deviation of the same types of workloads from the training data.  The accuracy
of the threshold value will be validated by the validation data.  For validation
step, a fingerprint data from the validation data set use its phase number to
identify first similar workload fingerprints from the training data and then
calculate ED and DTW with identified data. The smallest value of ED and DTW will
identify the most similar workload.  Then, we will check the threshold value of
particular workload to confirm.

\section{Evaluation}

\subsection{Experimental Setup}

\begin{figure}[!tbp]
\centering
\includegraphics[scale=0.46]{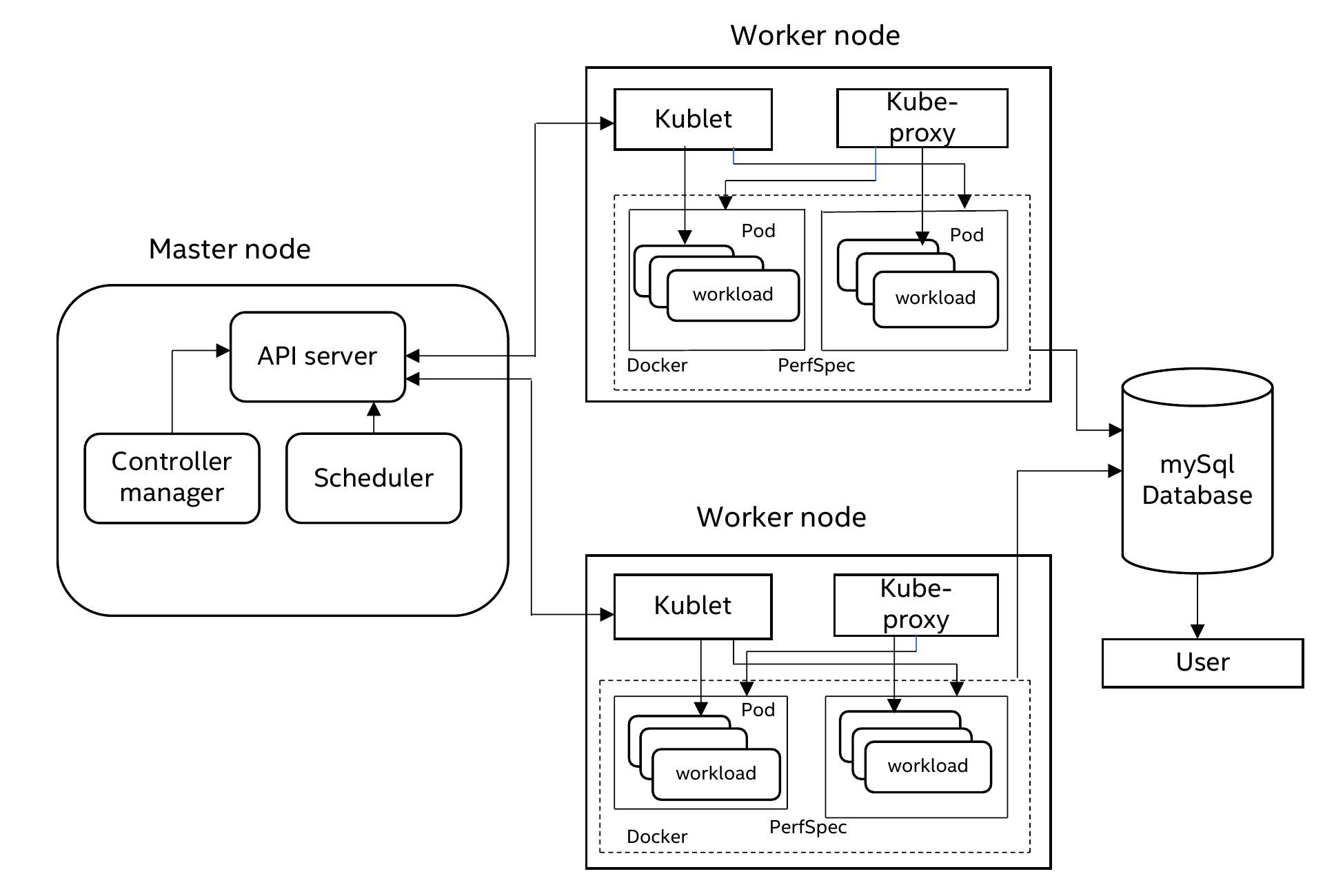}
\caption{System setup for data collection in Kubernets Cluster}
\label{kubernetes}
\end{figure}

To validate the proposed unknown workload prediction algorithm
and ascertain its accuracy, we set up a Kubernetes cluster, known as K8S.
Kubernetes is an open-source system for automating deployment, scaling and management of
containerized application. As shown in Figure~\ref{kubernetes} our K8s cluster
has 14 worker nodes and 1 master node.

\paragraph{Hardware setup}
All the worker nodes and the master node were running on separate Intel Xeon
processors (codenamed SkyLake), each having 2 sockets for a total of 28 cores, running
at a maximum frequency of 3.9GHz and 512GB of memory. All the servers were
running Ubuntu-20.04 server operating system.

\paragraph{Workloads} We ran a total of 6 training workloads in our experiments:
\texttt{mongoDB-ycsb-4.4}, \texttt{MySQL8}, \texttt{Wordpress-5MT},
\texttt{Bayesian inference}, \texttt{AI-benchmark-0.1.2} and \texttt{Redis-6.0.9}.
These workloads cover some of the typical applications run in cloud environment.
Each workload runs in a docker container.

\paragraph{Telemetry data collection}

We used a total of 72 hardware and software counters in our experiment (some of
them are shown in Table~\ref{table:hardware_and_software_counters}). We execute
the workloads for a total of 30 (CPU) seconds and sample the counters at every
second. We believe that the initial workload execution time contains different
phases such as warmup, connection setups, etc., which represent interesting set
of phases for experimentation. We use PerfSpect~\cite{Perf} tool
to collect hardware telemetry data. The PerfSpect tool, based on Linux
\texttt{perf} collects data from the underlying hardware platform, and it has
two main functions: 1) collect PMU (Performance Monitoring Unit) counters, and 2)
post-processing data and generate a CSV output of performance metrics.

\begin{table}[htbp]
\caption{Hardware and software counters}
\begin{center}
\begin{tabular}{l|l}
\hline \hline
\textbf{Counter} & \textbf{Type}  \\
\hline \hline
\texttt{Load Instructions} & Hardware \\
\hline
\texttt{LLC Cache Misses} & Hardware \\
\hline
\texttt{L1 cache misses} & Hardware \\
\hline
\texttt{iTLB misses} & Hardware \\
\hline
\texttt{L2 cache misses} & Hardware \\
\hline
\texttt{CPU utilization} & Software \\
\hline
\texttt{Page Major Faults} & Software \\
\hline
\texttt{Total Page Cache} & Software \\
\hline
\texttt{Canonical Memory Usage} & Software \\
\hline
\end{tabular}
\label{table:hardware_and_software_counters}
\end{center}
\end{table}

\subsection{Research Questions}

In our evaluation, we attempt to answer following research questions:
\begin{itemize}
\item \textbf{RQ1}: How accurately can we classify known workloads as similar?
\item \textbf{RQ2}: How does {\system} perform when classifying unknown
workloads?
\item \textbf{RQ3}: What could be a threshold for classfying similar
workloads?
\end{itemize}

We answer these questions by simulating 3 case studies.

We answer \textbf{RQ1} in the first case study by developing an ML based
supervised binary classification model that learns to classify workloads using
their fingerprints. Specifically, we utilize Gradient Boosting Tree (GBT)
algorithm to develop the prediction models due to the high dimensionality of the
workload fingerprints\footnote{We do not use neural network or deep
learning-based models since the numbers of sample runs is limited for each
workload.} We generate a labeled dataset by executing every workload 10 times
(and obtaining its 10 fingerprints).
For training the prediction model, we split the labeled dataset into
training and test set (70\%-30\%), and later use the test set to evaluate the
accuracy. Specifically, we split the dataset entries for every workload in
70\%-30\% manner; 7 entries per workload for training and 3 for evaluation.

We answer \textbf{RQ2} in the second case study by training the GBT-based
prediction model on the fingerprints of 4 workloads (MySQL, AI-benchmark, Redis and
WordPress) and evaluating the model's performance on 2 unknown workloads
(MongoDB-ycsb and Bayesian inference).

For \textbf{RQ3}, we obtain distances between multiple executions of every
workload from the set of 6 workloads and present those values
(Section~\ref{workload_similarity}).

\subsection{Results and Discussion}

\paragraph{\textbf{RQ1}}
Table~\ref{tab:prediction-output} summarizes the result for the first case
study. To summarize, the accuracy of GBT-based ML model is 93\%. Precision of
95\% additionally indicates lower false positive rate.

\begin{table}[!htbp]
  \centering
	\caption{Case Study 1: Accuracy of classifying known workloads}
	\label{tab:prediction-output}

    \begin{tabular}{l|l|l|l|l}
    \hline \hline
    \textbf{Name} & \textbf{Precision} & \textbf{Recall} & \textbf{F1 Score} & \textbf{Accuracy (\%)} \\
        \hline \hline
    GBT & 0.95 & 0.93 & 0.93 & 93 \\ \hline
    \end{tabular}
\end{table}

\paragraph{\textbf{RQ2}}

For case study 2, the results are presented in Table
\ref{tab:prediction-output02}.  The prediction model identified \texttt{Bayesian
inference} workload as similar to \texttt{AI-benchmark} workload 100\% of the
time. We speculate that this could be because \texttt{Bayesian inference}
workload uses Python-based framework for inference, which is also used by
\texttt{AI-benchmark} workload as it performs neural network-based inference. In
the case of \texttt{mongoDB-ycsb} workload, the prediction model identified it
as \texttt{WordPress} 67\% of the time and \texttt{MySQL} as 33\% of the time.
We speculate that this could be because of similar database operations performed
by all three workloads. Specifically, \texttt{WordPress} workload runs
pre-populated user blogging data with the \texttt{MariaDB} database, a fork of
MySQL.  Although, \texttt{MongoDB} is a NoSQL database, some of its general
database operations probably stress the hardware and software counters similar to
the \texttt{MySQL} workload and \texttt{WordPress}'s MariaDB operations.

\begin{table}[htbp]
\centering
\caption{Case Study 2: Classifying unknown workloads}
	\label{tab:prediction-output02}
\begin{tabular}{l|l|l|l|l} 
\hline \hline
\diagbox{\begin{tabular}[c]{@{}l@{}}\textbf{Unknown}
\\\textbf{workload}\end{tabular}}{\begin{tabular}[c]{@{}l@{}}\textbf{Predicted}
\\\textbf{workload}\end{tabular}} & \begin{tabular}[c]{@{}l@{}}~\textbf{AI-}\\\textbf{benchmark}\end{tabular} & \textbf{WordPress} & \textbf{MySQL} & \textbf{Redis}  \\ 
\hline \hline
\texttt{MongoDB-ycsb}                                                             & 0                                                       & 67\%      & 33\%  & 0      \\ 
\hline
\texttt{Bayesian Inference}                                                                                                                 & 100\%                                                   & 0         & 0     & 0      \\
\hline
\end{tabular}
\end{table}

\paragraph{\textbf{RQ3}}

\begin{figure}[!t]
\centering
\includegraphics[width=0.52\textwidth]{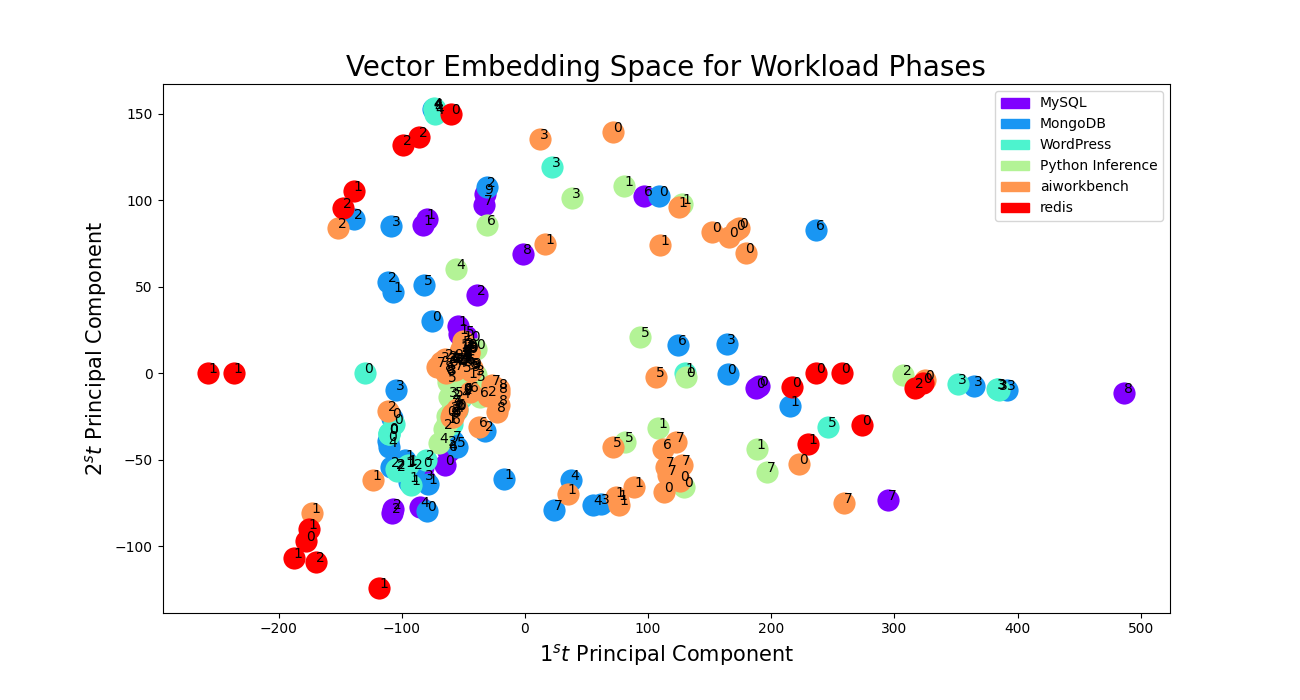}
\caption{Figure showing different workloads phases}
\label{visual}
\end{figure}

To answer the question about appropriate threshold values to classify similar
workloads, we run every workload multiple times and calculate the distances
between multiple runs of same workload.
Table~\ref{tab:phase-similarity} presents the mean and standard deviation 
of the distances for every workload.
In general, DTW-based distances are larger compared ED-based distances.  For both DTW and
ED based distances, the largest and the smallest values are observed in
\texttt{Redis} and \texttt{AI-benchmark} workloads respectively.
Intuitively, the mean distances represent radius of the workload-specific
circles, every point in which would represent a workload similar to that
workload. So we can use these mean values as the threshold values to classify
similar workloads. Figure~\ref{visual} shows the visual similarity of different
workload phases.


\begin{table}[!t]
\centering
\caption{Workload Similarity Measurement Using Fingerprints (DTW = dynamic time
warping; ED = Euclidean distance)}
\label{tab:phase-similarity}
\begin{tabular}{l||ll||ll}
\hline \hline
\textbf{Workload} & \multicolumn{2}{c||}{\textbf{DTW}} & \multicolumn{2}{c}{\textbf{ED}} \\ \hline
                  & \multicolumn{1}{l|}{\textbf{Mean}}  & \textbf{Standard} &
                    \multicolumn{1}{l|}{\textbf{Mean}}  & \textbf{Standard} \\
                  & \multicolumn{1}{l|}{} & \textbf{deviation} &
                  \multicolumn{1}{l|}{} & \textbf{deviation} \\ \hline \hline
\texttt{Redis}    & \multicolumn{1}{l|}{211.90} & 52.74 & \multicolumn{1}{l|}{28.06}   & 6.41 \\ \hline
\texttt{AI-benchmark}  & \multicolumn{1}{l|}{131.07} & 52.66 & \multicolumn{1}{l|}{19.93} & 7.33 \\ \hline
\texttt{MySQL}    & \multicolumn{1}{l|}{181.94} & 65.53 & \multicolumn{1}{l|}{23.51}   & 8.45 \\ \hline
\texttt{MongoDB-ycsb} & \multicolumn{1}{l|}{166.23} & 73.30 & \multicolumn{1}{l|}{23.19}  & 9.18 \\ \hline
\texttt{WordPress}    & \multicolumn{1}{l|}{145.25} & 64.47 & \multicolumn{1}{l|}{20.21}  & 6.94 \\ \hline
\texttt{Bayesian inference} & \multicolumn{1}{l|}{181.60} & 46.69 & \multicolumn{1}{l|}{27.93} & 6.32 \\ \hline
\end{tabular}
\end{table}

\section{Conclusion} 

In this paper, we present our system named {\system} that predicts workloads running in a
cloud environment. It profiles and characterizes cloud workload to understand
workload performance and phases. We classify cloud workloads using
software and hardware performance counters to generate workload-specific
fingerprints and develop formulas to measure similarity of workload phases.
Our experimental evaluation performed on a set of 6 cloud workloads demonstrates
high accuracy of our machine learning based model.

\section{Future work}

We plan to improve this work by increasing the number of workloads run in a
docker container so that we can improve workload prediction with a higher degree
of confidence for our future work. We also plan to implement in production so
that we can predict any unknown workload in real time.

\end{document}